\documentclass[twocolumn,letter]{jpsj2}
\title{Lattice Dynamics of LaFeAsO$_{1-x}$F$_x$ and PrFeAsO$_{1-y}$ \\
via Inelastic X-Ray Scattering and First-Principles Calculation}

\author{Tatsuo \textsc{Fukuda}$^{1,2}$\thanks{E-mail: fukuda@spring8.or.jp},
Alfred \textsc{Q.R. Baron}$^{1,3}$\thanks{E-mail: baron@spring8.or.jp},
Shin-ichi \textsc{Shamoto}$^{4}$, Motoyuki \textsc{Ishikado}$^{4}$,
Hiroki \textsc{Nakamura}$^{5}$, Masahiko \textsc{Machida}$^{5}$,
Hiroshi \textsc{Uchiyama}$^{1,3}$, Satoshi \textsc{Tsutsui}$^{3}$,
Akira \textsc{Iyo}$^{6}$, Hijiri \textsc{Kito}$^{6}$,
Jun'ichiro \textsc{Mizuki}$^{2}$, Masatoshi \textsc{Arai}$^{7}$,
Hiroshi \textsc{Eisaki}$^{6}$, Hideo \textsc{Hosono}$^{8}$}

\inst{$^{1}$Materials Dynamics Laboratory, SPring-8/RIKEN,
 Sayo, Hyogo 679-5148 \\
$^{2}$Synchrotron Radiation Research Unit, SPring-8/JAEA,
 Sayo, Hyogo 679-5148 \\
$^{3}$Research and Utilization Division, SPring-8/JASRI,
 Sayo, Hyogo 679-5198 \\
$^{4}$Neutron Materials Research Center, JAEA,
 Naka, Ibaraki 319-1195 \\
$^{5}$CREST, JST and the Center for Computational Science and e-Systems, JAEA,
 Taito, Tokyo 110-1195 \\
$^{6}$Nanoelectronics Research Institute (NeRI), AIST,
 Tsukuba, Ibaraki 305-8568 \\
$^{7}$J-PARC Center, JAEA, Naka, Ibaraki 319-1195 \\
$^{8}$ERATO-SORST, JST and Frontier Research Center, Tokyo Institute of Technology,
 Yokohama, 226-8503}

\abst{The lattice dynamics of LaFeAsO$_{1-x}$F$_x$ ($x$=0, 0.1) and
    PrFeAsO$_{1-y}$ ($y\sim$0.1) are investigated using inelastic x-ray
    scattering and {\it ab-initio} calculation.
    Measurements of powder samples provide an approximation to the
    phonon DOS, while dispersion is measured from a single crystal of
    PrFeAsO$_{1-y}$.
    A model that agrees reasonably well with all of the data at room
    temperature is built from results of {\it ab-initio} calculations by
    reducing the strength of the Fe-As bond by 30\%.}

\kword{phonon density of states, dispersion relation, softening,
iron-based superconductor, inelastic x-ray scattering, first-principles
calculation}

\begin{document}
\maketitle

The recent discovery of high-temperature superconductivity in
iron-arsenic materials,\cite{KamiharaY08a,TakahashiH08a,RotterM08a} with
$T_c$ exceeding 50K in some compounds,\cite{RenZA08a,KitoH08a} has led
to an outpouring of work reminiscent of that following the discovery of
the high $T_c$ of MgB$_2$.\cite{NagamatsuJ01a}
In the latter case, the synergy of experiment and calculation quickly
demonstrated that MgB$_2$ was an exotic phonon-mediated superconductor.
Despite the complexity of the new Fe-As compounds as compared to
MgB$_2$, calculations of phonon properties have appeared
quickly,\cite{SinghDJ08a,BoeriL08a} and seem to support the assertion
that these materials are not phonon mediated.\cite{BoeriL08a,HauleK08a}
However, these calculations require experimental verification.
Here we show that measurements of the phonon density of states (DOS) in
several samples, and phonon dispersion in a single crystal of
PrFeAsO$_{1-y}$ do not agree with LDA-based calculations, including
those referenced above.
However, the calculations can be brought into much closer agreement with
the data by softening the Fe-As bond.
This highlights the failure of the calculations to predict the
properties of the most important bond in this system, and suggests that
the results of those calculations, especially regarding phonon behavior,
should be carefully considered.

It is early to survey the properties of the FeAs superconductors, as
information is very quickly evolving.
However, the available phonon calculations based on pseudopotential
methods are essentially consistent with each
other,\cite{SinghDJ08a,BoeriL08a} and with our calculations here, and do
not indicate strong electron-phonon coupling.\cite{BoeriL08a}
Calculations using a rigid ion model also suggest small electron-phonon
coupling.\cite{HauleK08a}
The phonon DOS of a material provides a first, global, look at phonon
behavior.
In this case, it allows us to develop a model of the dynamics based on
the softening of just the Fe-As bond, which agrees reasonably with the
data.
The agreement is confirmed by comparison with spectra measured from a
single crystal of PrFeAsO$_{1-y}$.

All measurements were made using $\sim$1.5 meV resolution at 21.75 keV
at BL35XU\cite{BaronAQR00a} of SPring-8.
The LaFeAsO and LaFeAsO$_{0.9}$F$_{0.1}$ powder samples were synthesized
as discussed in ref. \citen{KamiharaY08a} and the PrFeAsO$_{1-y}$ was
prepared using a high-pressure growth as in ref. \citen{KitoH08a}.
X-ray diffraction showed the polycrystalline samples to be predominantly
single phase, and a careful four-circle investigation of the single
crystal, 150$\times$100$\times$20 $\mu m^3$, showed it to be a single
grain, with a mosaic spread of about 1$^\circ$.
$T_c$ of the samples were found to be 27K for the powder of
LaFeAsO$_{0.9}$F$_{0.1}$ and 48K for PrFeAsO$_{1-y}$, consistent with
refs. \citen{KamiharaY08a,KitoH08a} and 36K for the single crystal.
About 50 mg of each powder was gently pressed into a tube with a
diameter of 3 mm to form a flat surface for reflection-based inelastic
x-ray scattering measurements (they were confirmed to be un-oriented
after pressing) while the single crystal was mounted on a thin glass
capillary and measured in a Laue geometry.

Calculations of the phonon spectra of LaFeAsO were carried out using the
relaxed tetragonal {\it P4/nmm} structure from VASP\cite{KresseG93a}
calculated using a PAW method\cite{BlochlPE94a} and the GGA.
Phonons were then calculated via the PHONON package using a direct
method,\cite{ParlinskiK97a} in contrast to the perturbation methods of
refs. \citen{SinghDJ08a,BoeriL08a}.
In our calculations, the energy cut-off for plane waves was 550 eV, the
spacing of k-points was less than 0.1 {\AA}$^{-1}$ and the convergence
condition was that the total energy difference be less than 1 $\mu$eV.
The relaxed structural parameters were found to be $a$=4.0219 {\AA} and
$c$=8.6215 {\AA} with $z_{La}$=0.14496 and $z_{As}$=0.63817, similar to
refs. \citen{SinghDJ08a,BoeriL08a}.

\begin{figure}[tb]
\begin{center}
\includegraphics[keepaspectratio,width=0.35\textwidth]{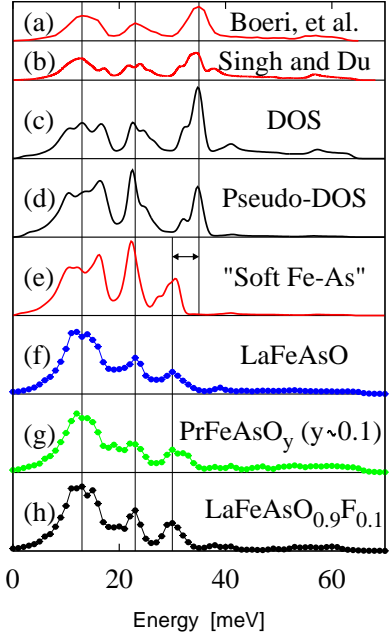}
\end{center}
\caption{Calculated and measured densities of states.
 The usual phonon density of states from (a) Boeri {\it et
 al.}\cite{BoeriL08a}, (b) Singh {\it et al.}\cite{SinghDJ08a}, and (c)
 this work.
 (d) The pseudo-DOS and (e) the effect of softening only the Fe-As bond
 (see discussion in text).
 (f)-(h) The pseudo-DOS extracted from measurements of three samples
 (see the Appendix for details on data processing).
 Vertical lines are at 13, 23, 30 and 35 meV, while the horizontal arrow
 shows the 5 meV shift between the room temperature measurements and the
 nominal calculation.
 The improved agreement with the soft Fe-As model is clear.}
\label{f1}
\end{figure}

Figure \ref{f1} shows both the results of our calculations, and the data
from several samples.
The upper panels show the agreement of our calculated phonon DOS with
previous work, and also the evolution of the DOS to the ``pseudo-DOS''
to be compared with experiment.
By pseudo-DOS we mean the calculated one-phonon x-ray scattering
integrated over the experimental range of momentum transfers.
In this case, we use a momentum full width of 3.7 nm$^{-1}$
(corresponding to the acceptance of the analyzer array) centered at a
momentum transfer of about 82 nm$^{-1}$ -- the precise value
(fluctuation $<$ 1 nm$^{-1}$) was sample dependent and was chosen to
reduce the elastic contribution due to Bragg peaks.
The DOS to pseudo-DOS transformation leaves the fine structure nearly
unaffected, while changing the over-all weighting of the peaks.
What is crucial for our discussion is that the triplet peak structure,
with a broad peak at 13 meV and narrower ones at 23 and 35 meV remains
robust, independent of which quantity is plotted.

The lower half of Fig. \ref{f1} compares the extracted pseudo-DOS (see
the appendix for details on data treatment) to the calculation.
For all compounds, the first and second peaks agree reasonably with the
calculation for LaFeAsO, however, the measured third peak is strongly
softened, about 5 meV or 15\%.
Looking more closely, one also notes that the first peak is narrower
experimentally than calculated, while the second and third peaks are
blurred out, and generally weaker than the calculated ones.
We also note the presence of a weak peak at about 39 meV in some of the
measured spectra, and also seen, at slightly higher energy, in the
calculation.
Overall, the clearest result is that the third peak in the DOS is
softened.
Considering the loss of definition of the peaks, however, one can
speculate that the phonon modes between 10 and 35 meV are somewhat
blurred out and soft, either due to band broadening (increased
dispersion), or, perhaps, lifetime effects.
Our measured DOS results are consistent with ref. \citen{QiuY08a} (for
$E<$20 meV) and unpublished
work.\cite{ChristiansonAD08a,HigashitaniguchiS08a}

\begin{figure}[tb]
\begin{center}
\includegraphics[keepaspectratio,width=0.5\textwidth]{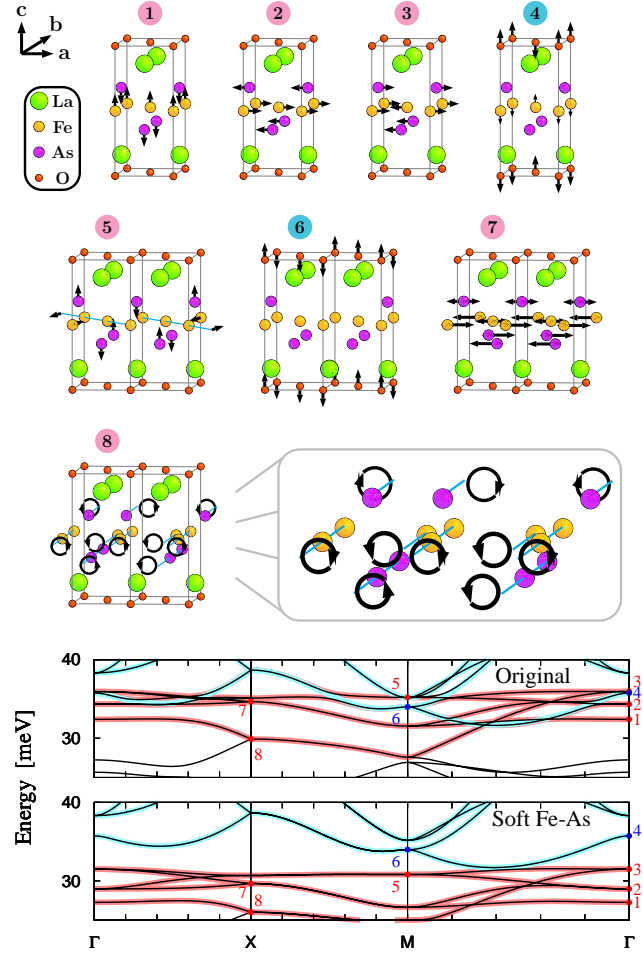}
\end{center}
\caption{Selected mode polarizations and dispersion.
 Polarizations are shown for the points indicated on the dispersion
 plots, with modes at M and X shown with two unit cells to allow the
 full phasing to be visualized.
 Modes at $\Gamma$ are generally either planar or $c$-axis polarized,
 while modes at M and X can be mixed.
 Modes at X, in particular, are complex and can involve circular
 motions.
 Red (blue) shading indicates Fe-As (oxygen) modes.
 The effect of the soft Fe-As bond is primarily to shift the Fe-As modes
 to lower energy, as compared to the original calculations.}
\label{f2}
\end{figure}

The dispersion and the contribution to the phonon modes near 35 meV
calculated peak appears somewhat complex (see Fig. \ref{f2} and also
discussion in ref. \citen{BoeriL08a}).
The relevant band of modes consists primarily of in-plane FeAs motion at
$\Gamma$, but the dispersion of these modes through the zone is
complicated by (1) mixing with the $c$-axis polarized motion of the FeAs
layers, and (2) the crossing of fast-dispersing oxygen modes.
However, the peak in the pseudo-DOS is due nearly entirely to Fe-As
motion.
This can be demonstrated by softening \textit{only} the Fe-As force
constant matrix in the calculations.
The resulting ``soft Fe-As'' pseudo-DOS calculation then agrees
reasonably well with the data: a reduction of the Fe-As nearest neighbor
force constant matrix to 70\% of the calculated value brings the
calculated peak into line with the measured one, and has only small
effect on the rest of the spectrum.

\begin{figure}[tb]
\begin{center}
\includegraphics[keepaspectratio,angle=-90,width=0.45\textwidth]{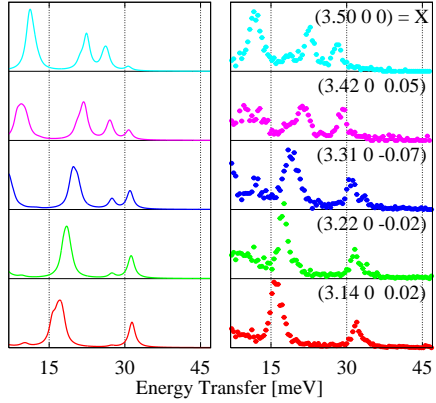}
\end{center}
\caption{Comparison of the measured dispersion for PrFeAsO$_{1-y}$
 ($y\sim$0.1), $T_c$=36K at room temperature (right panel) and
 calculation (left).
 See text for discussion.}
\label{f3}
\end{figure}

This model is in reasonable agreement with measurements of a small,
150$\times$100$\times$20 $\mu m^3$, single crystal of PrFeAsO$_{1-y}$
($y\sim$0.1), $T_c$=36K.
Figure \ref{f3} shows the dispersion along $\Gamma$-X in the (300)
Brillouin zone.
The high energy region shows the softening of the LO modes, essentially
as expected from our model, excepting near the transfer ($H$=3.31, 3.42)
of intensity from mode 3 in Fig. \ref{f2} (an in-plane polarized LO
mode) to mode 8 (a circular mode in the $a$-$c$ plane), where the data
shows a more abrupt transfer than the calculation.
The transition point is sensitive to small changes in polarization, and
could be affected by the oxygen nonstoichiometry.
The slightly complex structures at lower energy come from exciting
multiple nearby modes and are, partly, due to our finite momentum
resolution [$\Delta Q$= (0.067, 0.068, 0.035) rlu].
However, interestingly, the broad structure at about 10 meV at (3.42, 0,
0.05) is actually in better agreement with the original calculations
(not shown) and, will be the subject of future investigation.
The remaining small discrepancies in the lower-energy modes, given the
changes in system (La$\rightarrow$Pr), are not unreasonable.
We note that available Raman data\cite{HadjievVG08a,ZhaoSC08a} are in
better agreement with the soft Fe-As model than the original
calculation: the average energy difference between measurements and our
calculation is reduced from 1.9 to 0.9 meV (for three
modes)\cite{HadjievVG08a} and from 2.0 to 0.9 meV (for five
modes)\cite{ZhaoSC08a} when the softening is included in the model.

The softening of the phonons relative to the calculations suggests that
there is some character of the interaction that is not properly included
in the calculations, and this interaction screens the phonons and
reduces their energy.
This is consistent with the tendency the LDA calculations to
under-estimate the Fe-As bond distance (ie: 2.337 {\AA} in the present
calculation as compared to 2.407 {\AA} measured in ref.
\citen{delaCruzC08a}).
In this context we note that another large discrepancy between the
calculations and measured data regards the determination of the magnetic
moment on the iron atoms: the pseudopotential
calculations\cite{MazinII08a} give consistently much larger iron moments
(1 to 2 $\mu_B$) than measurements ($\sim$0.2-0.4
$\mu_B$)\cite{delaCruzC08a,KlaussHH08a,KitaoS08a}.
This has been, improved by going to an all-electron
calculation,\cite{YildirimT08a} but bond-lengths were not reported.
Also, a negative U\cite{NakamuraH08a} has been considered to account for
this, but appears not to strongly affect the phonon energies.
One pertinent fact, at least for the phonon question, regards the phase
transition in the parent material LaFeAsO: it transforms from a
tetragonal high temperature phase to an orthorhombic low temperature
phase at about 145 K, with magnetic order setting at $\sim$140K (see,
e.g., refs. \citen{delaCruzC08a} and \citen{KlaussHH08a}).
Carefully relaxed calculations suggest that the energy barrier between
the two possible orthorhombic configurations at low-temperature is very
small ($\sim$5 meV),\cite{SushkoPV08a} well within phonon energy scales.
Thus it is possible that anharmonicity plays a significant role in
determining material properties.
To investigate the effects of possible anharmonicity, we cooled the
LaFeAsO$_{0.9}$F$_{0.1}$ powder sample to 35K.
We found the low-energy peak at 13 meV hardened slightly, $<$ 1 meV, the
peak at 23 meV hardened about 1 meV, and the peak at 30 meV hardened
about 1.5 meV, similar to refs.
\citen{ChristiansonAD08a,HigashitaniguchiS08a}.
In order to get good agreement with calculation for the high-energy
peak, we still have to soften the Fe-As force constant matrix by about
20\%.
While this does not rule out an anharmonic potential leading to
softening of the Fe-As bond, it does clearly indicate that the strong
softening we observe is predominantly intrinsic, not related to the
room-temperature measurement.

In summary, our data shows that one peak in the phonon DOS at room
temperature is softened by about 15\% compared to LDA based calculations
of LaFeAsO, and that this softening can be well modeled by reducing just
the Fe-As nearest-neighbor force constant matrix by about 30\% from the
calculated value.
This model is consistent with single crystal measurements.
The softening of the peak in the pseudo-DOS is reduced to about 10\% at
35K, consistent with about 20\% reduction in the Fe-As force constant,
but is still large.
We also see a general blurring out of the spectral peaks that could be
consistent either with increased dispersion of the modes
(band-broadening) or possibly lifetime broadening, though the latter is
not likely based on the single crystal work.
These suggest rather strongly that the phonon system of this material is
not yet understood and that \textit{perhaps} one should not yet discard
a possible phonon mediated mechanism on the basis of the presently
available calculations.
If one were to assume the only effect of the softening on the Eliashberg
function, $\alpha^2$F, is to reduce some of the mode frequencies, then
indeed it would make negligible change to the calculated
$T_c$,\cite{BoeriL08a,ChristiansonAD08a} however, the source of the
softening is not clear, and, if it is due to coupling to the electronic
system (so changing the mode dependent electron-phonon coupling) then
the effect on calculated $T_c$ could be larger.
Notably, all samples investigated, including the non-superconducting
parent compound, show similar phonon DOS, suggesting that some other
change in, e.g., the electronic structure, switches on superconductivity
with doping.

\section*{Acknowledgment}
We thank K. Hashimoto, T. Shibauchi, R. Okazaki, and Y. Matsuda for
selection and characterization of the single crystal.
We are grateful to K. Yoshii for magnetization measurements on the
powder samples.
Work at SPring-8 was carried out under proposal numbers 2008A2050 and
2008A1981.
This work was performed under the NIMS-RIKEN-JAEA Cooperative Research
Program on Quantum Beam Science and Technology.

\appendix
\section{Data Analysis}

\begin{figure}[tb]
\begin{center}
\includegraphics[keepaspectratio,width=0.25\textwidth]{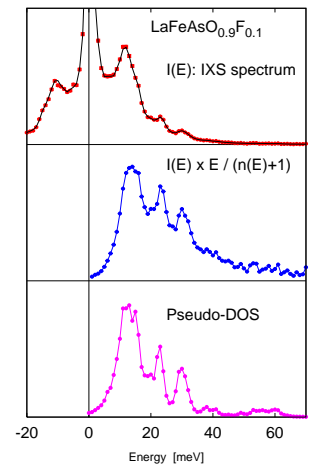}
\end{center}
\caption{Analysis progression for one data set at 300K.
 The top panel shows the measured data.
 The middle panel shows a scaled version, after subtraction of the
 elastic peak, that would approximate the GDOS in the limit of pure
 incoherent single-phonon scattering.
 The lowest panel shows the extracted pseudo-DOS after the multi-phonon
 contributions were removed by an iterative procedure.
 Note the solid line in the top panel is the back-calculated scattering
 from the pseudo-DOS, and its good agreement with the measurement
 demonstrates the self-consistency of the procedure.}
\label{fa1}
\end{figure}

The extraction of the ``pseudo-DOS'' from the measured scattering
proceeds much has been done previously in neutron scattering.
The procedure is exact in the case of pure incoherent scattering from a
monoatomic harmonic solid at a well-defined momentum transfer, but
approximate if there are significant other scattering processes, such
as, in the present case, coherent scattering.
A large range of integration over momentum space was chosen to attempt
to maximize the effective incoherent contribution and an iterative
fitting process was used to remove multiphonon contributions.
This provides an approximation to the generalized DOS (GDOS) which we
call the pseudo-DOS in the paper.
Early work along these lines can be found in ref. \citen{SjolanderA58a}
while a recent application to inelastic x-ray scattering, essentially
based on the analysis for nuclear resonant scattering, is discussed in
ref. \citen{BosakA05a}.
This treatment preserves most of the main features of the spectra, while
scaling them more nearly in agreement with the true one-phonon
scattering (see Fig. \ref{fa1}).
One should note that the extracted spectra at low energies are sensitive
to the subtraction of the elastic peak, so that results below 8 meV
should be taken as approximate.

\begin{thebibliography}{99} 
\bibitem{KamiharaY08a} Y. Kamihara, T. Watanabe, M. Hirano, and
	H. Hosono: J. Am. Chem. Soc. \textbf{130} (2008) 3296.
\bibitem{TakahashiH08a}  H. Takahashi, K. Igawa, K. Arii, Y. Kamihara,
	M. Hirano, and H. Hosono: Nature \textbf{453} (2008) 376.
\bibitem{RotterM08a} M. Rotter, M. Tegel, and D. Johrendt: arXiv:0805.4630.
\bibitem{RenZA08a} Z.-A. Ren, W. Lu, J. Yang, W. Yi, X.-L. Shen,
	Z.-C. Li, G.-C. Che, X.-L. Dong, L.-L. Sun, F. Zhou, and
	Z.-X. Zhao: Chinese Phys. Lett. \textbf{25} (2008) 2215.
\bibitem{KitoH08a} H. Kito, H. Eisaki, and A. Iyo:
	J. Phys. Soc. Jpn. \textbf{77} (2008) 063707.
\bibitem{NagamatsuJ01a} J. Nagamatsu, N. Nakagawa, T. Muranaka,
	Y. Zenitani, and J. Akimitsu: Nature \textbf{410} (2001) 63.
\bibitem{SinghDJ08a} D.J. Singh and M.-H. Du:
	Phys. Rev. Lett. \textbf{100} (2008) 237003.
\bibitem{BoeriL08a} L. Boeri, O.V. Dolgov, and A.A. Golubov:
	Phys. Rev. Lett. \textbf{101} (2008) 026403.
\bibitem{HauleK08a} K. Haule, J.H. Shim, and G. Kotliar:
	Phys. Rev. Lett. \textbf{100} (2008) 226402.
\bibitem{BaronAQR00a} A.Q.R. Baron, Y. Tanaka, S. Goto, K. Takeshita,
	T. Matsushita, and T. Ishikawa: J. Phys. Chem. Solids
	\textbf{61} (2000) 461.
\bibitem{KresseG93a} G. Kresse and J. Hafner: Phys Rev. B \textbf{47}
	(1993) 558(R); G. Kresse and J. Furthm\"{u}ller:
	Comput. Mat. Sci. \textbf{6} (1996) 15; G. Kresse and
	J. Furthm\"{u}ller: Phys. Rev. B \textbf{54} (1996) 11169.
\bibitem{BlochlPE94a} P.E. Bl\"{o}chl: Phys. Rev. B \textbf{50} (1994)
	17953; G. Kresse and D. Joubert: Phys. Rev. B \textbf{59} (1999)
	1758.
\bibitem{ParlinskiK97a} K. Parlinski, Z.Q. Li, and Y. Kawazoe:
	Phys. Rev. Lett. \textbf{78} (1997) 4063.
\bibitem{QiuY08a} Y. Qiu, M. Kofu, W. Bao, S.-H. Lee, Q. Huang,
	T. Yildirim, J.R.D. Copley, J.W. Lynn, T. Wu, G. Wu, and
	X.H. Chen: Phys. Rev. B \textbf{78} (2008) 052508.
\bibitem{ChristiansonAD08a} A.D. Christianson, M.D. Lumsden, O. Delaire,
	M.B. Stone, D.L. Abernathy, M.A. McGuire, A.S. Sefat, R. Jin,
	B.C. Sales, D. Mandrus, E.D. Mun, P.C. Canfield, J.Y.Y. Lin,
	M. Lucas, M. Kresch, J.B. Keith, B. Fultz, E.A. Goremychkin, and
	R.J. McQueeney: arXiv:0807.3370.
\bibitem{HigashitaniguchiS08a} S. Higashitaniguchi, M. Seto, S. Kitao,
	Y. Kobayashi, M. Saito, R. Masuda, T. Mitsui, Y. Yoda,
	Y. Kamihara, M. Hirano, and H. Hosono: arXiv:0807.3968.
\bibitem{HadjievVG08a} V.G. Hadjiev, M.N. Iliev, K. Sasmal, Y.-Y. Sun,
	and C.W. Chu: Phys. Rev. B \textbf{77} (2008) 220505(R).
\bibitem{ZhaoSC08a} S.C. Zhao, D. Hou, Y. Wu, T.L. Xia, A.M. Zhang,
	G.F. Chen, J.L. Luo, N.L. Wang, J.H. Wei, Z.Y. Lu, and
	Q.M. Zhang: arXiv:0806.0885.
\bibitem{delaCruzC08a} C. de la Cruz, Q. Huang, J.W. Lynn, J. Li,
	W. Ratcliff II, J.L. Zarestky, H.A. Mook, G.F. Chen, J.L. Luo,
	N.L. Wang, and P. Dai: Nature \textbf{453} (2008) 899.
\bibitem{MazinII08a} See I.I. Mazin, M.D. Johannes, L. Boeri,
	K. Koepernik, and D.J. Singh: arXiv:0806.1869 and references
	therein.
\bibitem{KlaussHH08a} H.-H. Klauss, H. Luetkens, R. Klingeler, C. Hess,
	F.J. Litterst, M. Kraken, M.M. Korshunov, I. Eremin,
	S.-L. Drechsler, R. Khasanov, A. Amato, J. Hammann-Borrero,
	N. Leps, A. Kondrat, G. Behr, J. Werner, and B. B\"{u}chner:
	Phys. Rev. Lett. \textbf{101} (2008) 077005.
\bibitem{KitaoS08a} S. Kitao, Y. Kobayashi, S. Higashitaniguchi,
	M. Saito, Y. Kamihara, M. Hirano, T. Mitsui, H. Hosono, and
	M. Seto: arXiv:0805.0041.
\bibitem{YildirimT08a} T. Yildrim: Phys. Rev. Lett. \textbf{101} (2008)
	057010.
\bibitem{NakamuraH08a} H. Nakamura, N. Hayashi, N. Nakai, and
	M. Machida: arXiv:0806.4804.
\bibitem{SushkoPV08a} P.V. Sushko, A.L. Shluger, M. Hirano, and
	H. Hosono: arXiv:0807.2213.
\bibitem{SjolanderA58a} See. e.g., A. Sj\"{o}lander: Arkiv f\"{u}r Fys.
	\textbf{14} (1958) 315; D.T. Keating: J. Phys. Chem. Solids
	\textbf{31} (1970) 1317.
\bibitem{BosakA05a} A. Bosak and M. Krisch: Phys. Rev. B \textbf{72}
	(2005) 224305.
\end{thebibliography}
\end{document}